\documentclass[twocolumn,secnumarabic,amssymb, nobibnotes, aps, prd]{revtex4-1}

\usepackage{color}
\usepackage{threeparttable}
\usepackage{srcltx}

\usepackage{graphicx,amsmath,psfrag,bm}
\usepackage[usenames,dvipsnames]{xcolor}
\usepackage{subfig}
\usepackage{cancel}
\usepackage{epstopdf}
\usepackage{pstool}

\newcommand{\ve}[1]{\boldsymbol{#1}}
\newcommand{\te}[1]{\overline{\overline{#1}}}

\begin{document}

\title{Space-Wave Routing via Surface Waves Using a Metasurface System}
\author{Karim~Achouri, and~Christophe~Caloz}

\email[E-mail: ]{karim.achouri@polymtl.ca, christophe.caloz@polymtl.ca}
\affiliation{Department of Electrical Engineering, Polytechnique Montr\'{e}al, Montr\'{e}al, Quebec, Canada}

\begin{abstract}
We introduce the concept of a metasurface system able to route space wave via surface waves. This concept may be used to laterally shift or modulate the beam width of scattered waves. We propose two corresponding synthesis techniques, one that is exact but leads to practically challenging material parameters and one that is approximate but leads to simpler material parameters. The concept is experimentally verified in an electromagnetic periscope. Additionally, we propose two other potential applications namely a beam expander and a multi-wave refractor.
\end{abstract}

\maketitle

\section{Introduction}

Metasurfaces are thin electromagnetic films composed of flat scatterers and represent the two-dimensional counterparts of volume metamaterials~\cite{kildishev2013planar,yu2014flat}. Over recent years, they have attracted tremendous attention due to their unprecedented capabilities to control electromagnetic waves conjugated with their ease of fabrication, low loss and high compactness.

The vast majority of metasurface designs and applications reported to date have been restricted to isolated metasurfaces, i.e. single metasurface structures performing specific electromagnetic transformations. In order to extend the range of these transformations, we propose here the concept of a \emph{metasurface system}, namely a combination of several metasurface structures collectively exhibiting properties that would be unattainable with a single metasurface. Specifically, we present a metasurface system composed of three juxtaposed metasurfaces, that \emph{routes} space-wave beams, between different locations, via surface waves. Such a system may be used, for instance, to laterally shift or modulate the beam width of scattered waves.

This paper is organized as follows. Section~\ref{sec:2} introduces the concept of space-wave routing via surface waves in a metasurface system. Then, Sec.~\ref{sec:3} discusses two synthesis techniques for the design of such a system. Based on the second proposed synthesis method, we experimentally demonstrate system routing in an ``electromagnetic periscope,'' whose design and measurements are presented in Sec.~\ref{sec:4}. Finally, Sec.~\ref{sec:5} proposes two additional potential applications, namely a compact beam expander and a multi-wave refractor. Conclusions are given in Sec.~\ref{sec:6}.

\section{Space-Wave Routing Concept}
\label{sec:2}

\begin{figure}[htbp]
\centering
\includegraphics[width=1\linewidth]{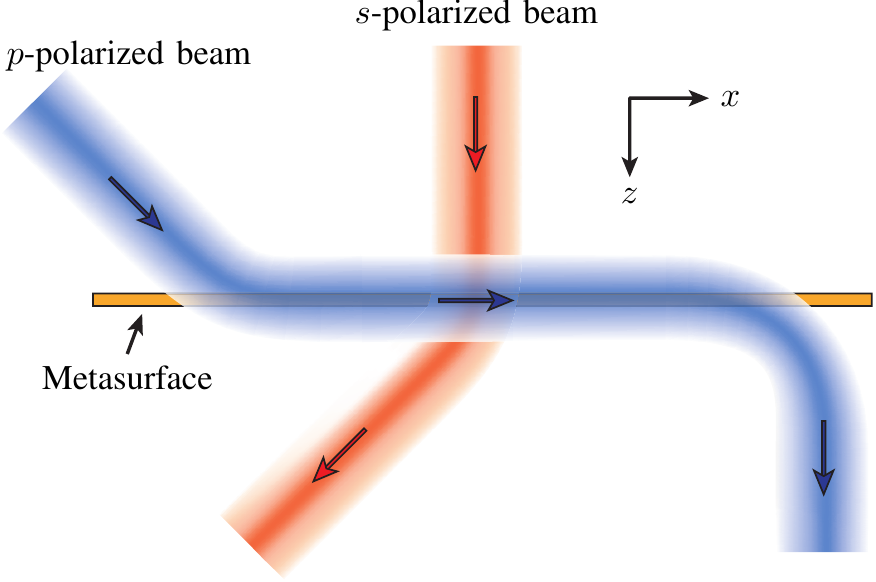}
%\psfragfig*[width=\linewidth]{SS_wave_concept}{
%\psfrag{a}[][][1]{$s$-polarized beam}
%\psfrag{b}[][][1]{$p$-polarized beam}
%\psfrag{c}[][][1]{Metasurface}
%\psfrag{x}[][][1]{$x$}
%\psfrag{z}[][][1]{$z$}}
\caption{Concept of metasurface system performing the operations of space-wave routing via surface waves for $p$-polarized beams and generalized refraction for $s$-polarized beams.}
\label{Fig:concept}
\end{figure}

The fundamental idea, which is depicted in Fig.~\ref{Fig:concept}, consists in converting an incoming space wave into a surface wave, propagating this surface wave between two points along a desired path, and then converting it back, with possible other transformations, into an outgoing space wave. This concept may be used to laterally shift reflected or transmitted waves (electromagnetic periscope), modulate the width of beams, or enable multiple refraction, in a very compact fashions, as will be discussed thereafter.
In the system depicted in Fig.~\ref{Fig:concept}, the metasurface is assumed to be monoanisotropic diagonal, and hence birefringent, allowing for the independent control of $s$ and $p$ polarizations. The metasurface may be designed, for instance, to route $p$-polarized waves and refract (or perform any another transformation on) $s$-polarized waves.

\begin{figure}[htbp]
\centering
\subfloat[]{\label{Fig:Compa1}
\includegraphics[width=1\linewidth]{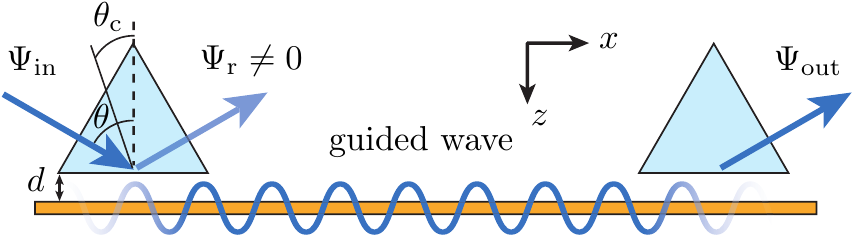}
%\psfragfig*[width=\linewidth]{Prism}{
%\psfrag{a}[][][1]{$\Psi_\text{r} \neq 0$}
%\psfrag{b}[][][1]{guided wave}
%\psfrag{c}[][][1]{$\Psi_\text{out}$}
%\psfrag{d}[][][1]{$d$}
%\psfrag{e}[][][1]{$\Psi_\text{in}$}
%\psfrag{f}[][][1]{$\theta$}
%\psfrag{g}[][][1]{$\theta_\text{c}$}
%\psfrag{x}[][][1]{$x$}
%\psfrag{z}[][][1]{$z$}}
}\\
\subfloat[]{\label{Fig:Compa2}
\includegraphics[width=1\linewidth]{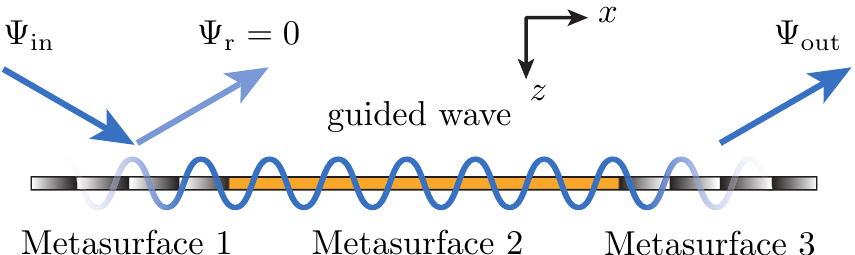}
%\psfragfig*[width=\linewidth]{Meta}{
%\psfrag{a}[][][1]{$\Psi_\text{r} = 0$}
%\psfrag{b}[][][1]{guided wave}
%\psfrag{c}[][][1]{$\Psi_\text{out}$}
%\psfrag{e}[][][1]{$\Psi_\text{in}$}
%\psfrag{m}[][][1]{Metasurface 1}
%\psfrag{n}[][][1]{Metasurface 2}
%\psfrag{p}[][][1]{Metasurface 3}
%\psfrag{x}[][][1]{$x$}
%\psfrag{z}[][][1]{$z$}}
}
\caption{Representations of two optical systems performing the same wave routing operation. (a)~Combination of two prisms and a dielectric waveguide. (b)~Composite metasurface, including two spatially modulated metasurfaces placed at the ends of a guiding metasurface.}
\label{Fig:Compa}
\end{figure}

We shall now describe the space-wave routing concept in more details. Let us consider the optical system depicted in Fig.~\ref{Fig:Compa1}, which consists of a dielectric waveguide with two prisms placed at different locations above it. This system may be used to perform the routing operation described in Fig.~\ref{Fig:concept}. Assume that an input beam $\Psi_\text{in}$ is impinging on the left prism at an angle $\theta > \theta_\text{c}$, where $\theta_\text{c}$ is the angle of total internal reflection. An evanescent wave with wavenumber $k_x$, corresponding to that of the incident wave, is formed between the prism and the waveguide due to total internal reflection. This evanescent wave then couples to a waveguide mode with matched $k_x$, and the resulting wave propagates along the waveguide in the $+x$-direction. The amount of coupling between the incident space wave and the guided wave is proportional to the distance $d$ between the prism and the waveguide, and is usually less than unity, leading to a non-zero reflected wave $\Psi_\text{r}$. Farther along the waveguide, the guided wave is transformed back into an output space wave $\Psi_\text{out}$ by the second prism by the reverse mechanism.

We introduce here the metasurface system depicted in Fig.~\ref{Fig:Compa2} to perform the same operation in a much more compact (purely planar) and (ideally) perfectly reflection-less fashion. This system consists of three different metasurfaces juxtaposed to each other. The input space wave is coupled into a guided surface wave by a spatially modulated metasurface. The middle metasurface is a surface-wave guiding structure that propagates the guided wave in the $+x$-direction. Finally, another spatially modulated metasurface transforms the guided wave back into a space wave at the other end of the system.

\section{Metasurface System Synthesis}
\label{sec:3}
\subsection{Exact Synthesis Based on GSTCs}
\label{sec:Ex_Synthesis_GSTCs}

The metasurface system introduced above can be rigorously synthesized so as to provide the exact medium parameters performing the transformation depicted in Fig.~\ref{Fig:Compa2}. For mathematical convenience, the metasurface is assumed to be of zero thickness. This assumption allows one to describe it as a spatial electromagnetic discontinuity, for which rigorous continuity conditions, the generalized sheet transition conditions (GSTCs), are available~\cite{Idemen1973,kuester2003av}. A complete presentation of this synthesis method was given in~\cite{achouri2014general,AchouriEPJAM}.

In the case of a monoanisotropic metasurface lying in the $xy$-plane at $z=0$, the GSTCs~\footnote{The time dependence $e^{j\omega t}$ is omitted throughout the paper.} read
\begin{subequations}
\label{eq:BC_plane}
\begin{align}
\hat{z}\times\Delta\ve{H}
&=j\omega\epsilon_0\te{\chi}_\text{ee}\ve{E}_\text{av},\label{eq:BC_plane_1}\\
\Delta\ve{E}\times\hat{z}
&=j\omega\mu_0 \te{\chi}_\text{mm}\ve{H}_\text{av},\label{eq:BC_plane_2}
\end{align}
\end{subequations}
where $\te{\chi}_\text{ee}$ and $\te{\chi}_\text{mm}$ are the electric and magnetic susceptibilities of the metasurface, respectively, $\Delta\ve{E}$ and $\Delta\ve{H}$ are the difference of the electric and magnetic fields on both sides of the metasurface, and $\ve{E}_\text{av}$ and $\ve{H}_\text{av}$ are the arithmetic averages of these fields. In the problem considered here, no rotation of polarization is required and therefore the monoanisotropic susceptibility tensors are purely diagonal. Moreover, it is assumed that the metasurface is not polarizable in its longitudinal direction which would otherwise lead to more complicated GSTC relations than~\eqref{eq:BC_plane}~\cite{achouri2014general}. Solving~\eqref{eq:BC_plane} to express the susceptibilities as a function of the specified fields yields the closed-form relations
\begin{subequations}
\label{eq:chi_diag}
\begin{align}
\chi_{\text{ee}}^{xx}&=\frac{-\Delta H_{y}}{j\omega\epsilon_0  E_{x,\text{av}}},\quad \chi_{\text{mm}}^{yy}=\frac{-\Delta E_{x}}{j\omega\mu_0  H_{y,\text{av}}},\label{eq:chi_diag1}\\
\chi_{\text{ee}}^{yy}&=\frac{\Delta H_{x}}{j\omega\epsilon_0  E_{y,\text{av}}},\quad \chi_{\text{mm}}^{xx}=\frac{\Delta E_{y}}{j\omega\mu_0  H_{x,\text{av}}},\label{eq:chi_diag2}
\end{align}
\end{subequations}
which describe a birefringent metasurface able to independently control and transform $x-$ and $y-$polarized waves.

Let us now synthesize the space-wave to surface-wave transformation performed by the first metasurface in Fig.~\ref{Fig:Compa2}. Let us assume a $p$-polarized wave ($\ve{E}\in xz$-plane and $\ve{H}\parallel\hat{y}$), to be routed (Fig.~\ref{Fig:concept}), which corresponds to the synthesis relations~\eqref{eq:chi_diag1}. In this case, the tangential electromagnetic fields, at $z=0$, are
\begin{equation}
\label{eq:EH}
E_x^{\text{a}}=A^{\text{a}}\frac{k_z^{\text{a}}}{k_0}e^{-jk_x^{\text{a}}x}\quad \text{and} \quad
H_y^{\text{a}}=A^{\text{a}}e^{-jk_x^{\text{a}}x}/\eta_0,
\end{equation}
where $A$ is a complex constant, $k_x$ and $k_z$ are the tangential and longitudinal wavenumbers, respectively, $\eta_0$ and $k_0$ are the impedance and wavenumber of free-space, respectively, and the superscript a $=$ i, r, t denotes the incident, reflected and transmitted waves, respectively. We shall consider the transformation of an incident space wave with $A^\text{i}=1$ and $k_x^\text{i}=k_0\sin{\theta^\text{i}}$, where $\theta^\text{i}=45^\circ$ is the incidence angle, into a surface wave with $A^\text{t}=0.728$ and $k_x^\text{t}=1.2k_0$. The corresponding longitudinal $k$-component is found as $k_z^\text{a}=\sqrt{k_0^2 - (k_x^\text{a})^2}$ which, in the case of the transmitted wave, is an imaginary quantity corresponding to wave evanescence perpendicular to the metasurface and surface-wave propagation in the $+x$-direction. The value 0.728 was derived to ensure a purely passive (although lossy) reflection-less ($A^\text{r}=0$) metasurface~\footnote{This proviso was found by inserting~\eqref{eq:EH} into~\eqref{eq:chi_diag1} with specified parameters $A^\text{i}$, $k_x^\text{a}$ and $k_z^\text{a}$ and solving for $A^\text{t}$ such that $\Im(\chi_{\text{ee}}^{xx}),\Im(\chi_{\text{mm}}^{yy})<0$. This shows the space wave cannot be transformed into a surface wave without dissipation.}.

\begin{figure}[htbp]
\centering
\subfloat[]{
\includegraphics[width=0.9\linewidth]{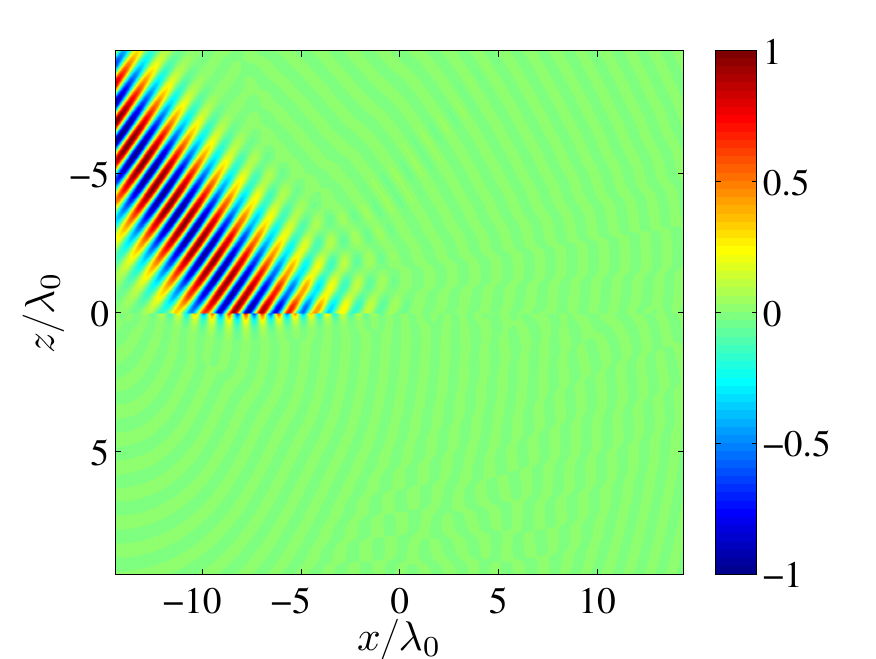}
%\psfragfig*[width=\linewidth]{1}{
%\psfrag{x}[][][1.2]{$x/\lambda_0$}
%\psfrag{z}[][][1.2]{$z/\lambda_0$}}
\label{Fig:Sim1}}\\
\subfloat[]{
\includegraphics[width=0.9\linewidth]{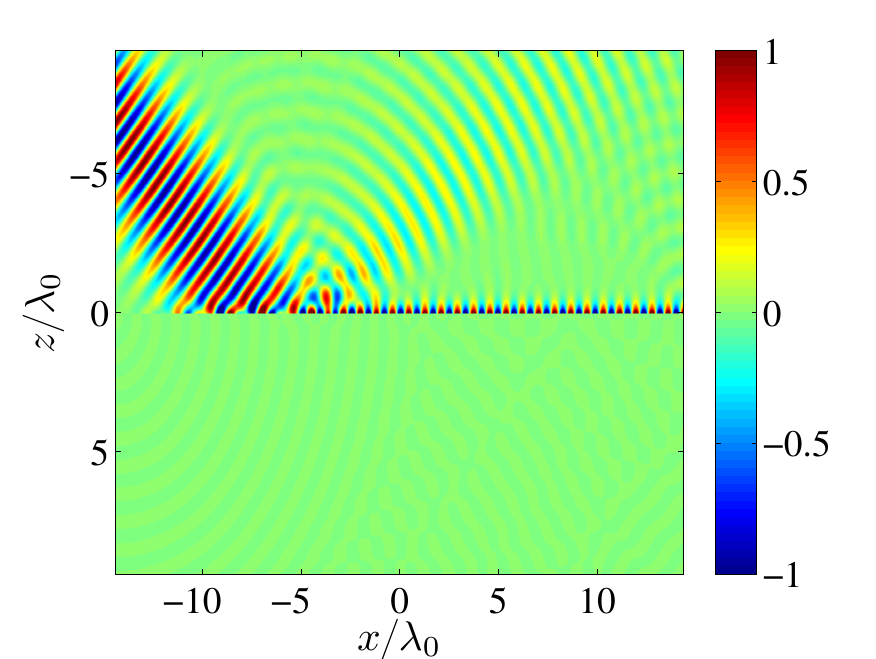}
%\psfragfig*[width=\linewidth]{2}{
%\psfrag{x}[][][1.2]{$x/\lambda_0$}
%\psfrag{z}[][][1.2]{$z/\lambda_0$}}
\label{Fig:Sim2}}\\
\subfloat[]{
\includegraphics[width=0.9\linewidth]{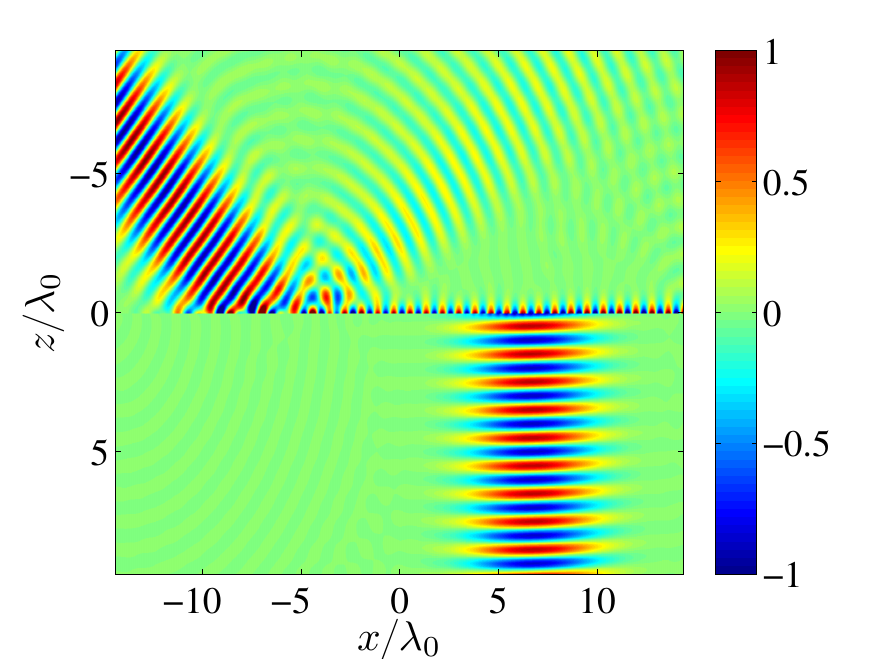}
%\psfragfig*[width=\linewidth]{3}{
%\psfrag{x}[][][1.2]{$x/\lambda_0$}
%\psfrag{z}[][][1.2]{$z/\lambda_0$}}
\label{Fig:Sim3}}
\caption{Finite-difference frequency-domain (FDFD) simulations showing the real part of $H_y$ in the case of (a)~the conversion of a space wave into a localized surface wave, (b)~the coupling of the surface wave into a guided wave that propagates along a juxtaposed metasurface, and (c)~the propagating surface wave is then transformed back into a space wave.}
\label{Fig:Sim}
\end{figure}

Finite-difference frequency domain (FDFD) simulations~\cite{vahabzadeh2016simulation} are used to analyse the response of the synthesized metasurface. Figure~\ref{Fig:Sim1} shows how an obliquely incident Gaussian beam is transformed into a surface wave on the transmit side of the metasurface. Note that the transformation in Fig.~\ref{Fig:Sim1} is ``perfect'' in the sense that no parasitic diffraction order is present. However, the surface wave only exists in the region where the Gaussian beam illuminates the metasurface and does not propagate farther along the structure. This is because the metasurface was synthesized assuming an incident plane wave illuminating the entire structure and not just a small portion of it, as is the case with a Gaussian beam. Consequently, the surface wave is restricted to the region within the waist of the incident beam and cannot propagate beyond its excitation zone as it is not an eigen-mode of this metasurface.

In order to propagate the surface wave farther along the surface, it is necessary to introduce a discontinuity in the metasurface or, in other words, to place a second metasurface next to the first one, as shown in Fig.~\ref{Fig:Compa2}. Thus, the first metasurface is synthesized as a space-wave to surface-wave transformer to convert the incident wave into a guided wave, while the second metasurface is synthesized as a surface-wave guiding structure, to route the wave along the overall structure. This operation is achieved, considering the juxtaposed (first and second) metasurface system as a composite metasurface, by specifying the fields as follows. The incident field is defined with the parameters $A^\text{i}=H(-x-5\lambda_0)$ and $k_x^\text{i}=k_0\sin{\theta^\text{i}}$ ($\theta^\text{i}=45^\circ$), where $H(x)$ is the Heaviside function. The Heaviside function is used here to create a discontinuity in the incident field at the position $-5\lambda_0$ on the metasurface. Additionally, we set $A^\text{t}=0$ and $A^\text{r}=0.728$ and $k_x^\text{r}=1.2k_0$~\footnote{In this second example, the surface wave is placed on the reflection side of the metasurface to later permit an easier design of the third metasurface.}. Inserting these field specifications into~\eqref{eq:chi_diag1} and performing an FDFD simulation for an incident Gaussian beam impinging on the metasurface at the position $-7\lambda_0$ yields the result presented in Fig.~\ref{Fig:Sim2}. As can be seen, the surface wave effectively couples into the second metasurface where it now propagates as a surface wave. Note that the presence of the discontinuity between the two metasurfaces introduces some spurious scattering of the incident wave, which could be avoided using a smooth transition.

Finally, the energy carried by the surface wave is extracted and transformed back into a space wave by the third metasurface in Fig.~\ref{Fig:Compa2} upon specifying a non-zero transmitted wave with parameters \mbox{$A^\text{t}=0.6\cdot\text{exp}[4(x-6.5\lambda_0)^2/5]$} and $k_x^\text{t}=k_0\sin{\theta^\text{t}}$, where the transmission angle is chosen here to be $\theta^\text{t}=0$. The simulation result is shown in Fig.~\ref{Fig:Sim3}.

To synthesize the second metasurface for refraction of the $s$-polarized wave, as shown in Fig.~\ref{Fig:concept}, one would simply need to insert the $s$-counterpart of~\eqref{eq:EH} into~\eqref{eq:chi_diag2}, as conventionally done for generalized refractive metasurfaces~\cite{achouri2014general}. In this case, assuming a non-zero transmission angle, the metasurface becomes globally nonuniform in the $x$-direction, although it is seen as perfectly uniform to the $p$-polarized wave (birefringence).

\begin{figure}[htbp]
\centering
\subfloat[]{
\includegraphics[width=1\linewidth]{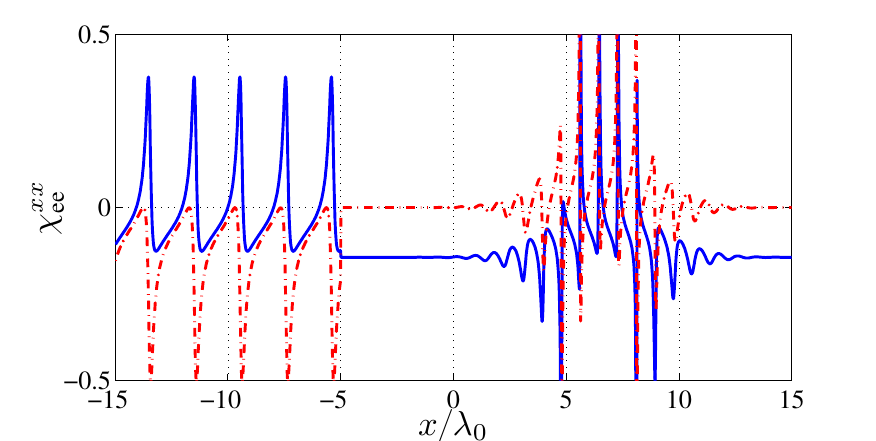}
%\psfragfig*[width=\linewidth]{Xe}{
%\psfrag{x}[][][1]{$x/\lambda_0$}
%\psfrag{y}[][][1]{$\chi_\text{ee}^{xx}$}}
\label{Fig:Xs1}}\\
\subfloat[]{
\includegraphics[width=1\linewidth]{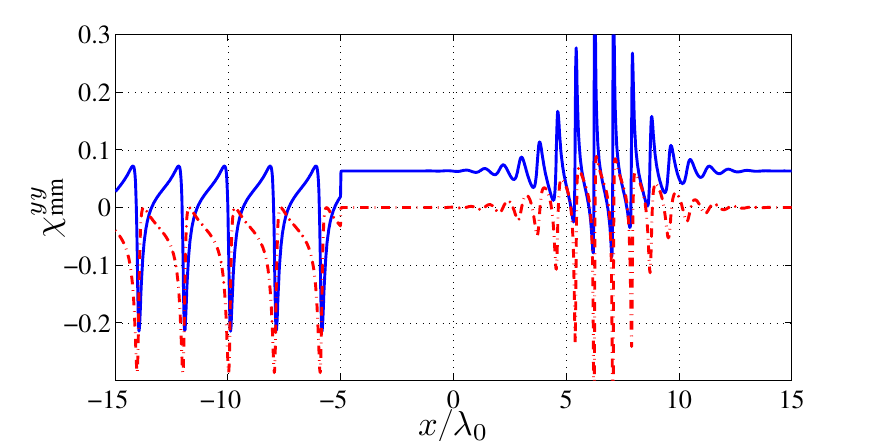}
%\psfragfig*[width=\linewidth]{Xm}{
%\psfrag{x}[][][1]{$x/\lambda_0$}
%\psfrag{y}[][][1]{$\chi_\text{mm}^{yy}$}}
\label{Fig:Xs2}}
\caption{Electric (a) and magnetic (b) susceptibilities corresponding to the transformation presented in Fig.~\ref{Fig:Sim3}. The solid blue lines are the real parts while the dashed red lines are the imaginary parts.}
\label{Fig:Xs}
\end{figure}

The electric and magnetic susceptibilities of the composite metasurface corresponding to the simulation in Figs.~\ref{Fig:Sim3}, are plotted in Fig.~\ref{Fig:Xs1} and~\ref{Fig:Xs2}, respectively. The conversion from space wave to surface wave occurs in the portion of the metasurface where $x<-5\lambda_0$. In this region, the metasurface (first metasurface) is spatially varying and exhibits nonuniform loss as is evidenced by the oscillating negative imaginary parts of the susceptibilities. From \mbox{$x=-5\lambda_0$} to approximatively $x=0$, the metasurface (second metasurface) supports the propagation of a surface wave. It is interesting to note that, in this region, the metasurface is perfectly uniform, passive and lossless, with susceptibilities given by the following simple relations $\chi_\text{ee}^{xx}=2j/k_z^\text{r}$ and $\chi_\text{mm}^{yy}=2jk_z^\text{r}$, where $k_z^\text{r}$ is the purely imaginary propagation constant of the surface wave in the longitudinal direction.
Finally, starting from $x>0$, the metasurface (third metasurface) becomes spatially varying again, allowing part of the energy conveyed by the surface wave to progressively leak out to form the space wave. Here, the susceptibilities have values oscillating between positive and negative imaginary parts. This indicates that the metasurface is successively varying between gain and loss. The presence of loss, as in the first part of the metasurface, is generally required to suppress undesired diffraction orders. The presence of active regions, corresponding to gain in the last part of the metasurface, is due to the way the fields were specified in the synthesis. Indeed, the surface wave (reflected wave) was specified with \emph{constant} amplitude over the entire metasurface, including in the third region, and it is therefore not surprising that gain is required in the region where the transmitted space wave is generated and where it draws power from the surface wave. The metasurface could be made perfectly passive by specifying a surface wave with progressively decreasing amplitude as its energy is being leaked out. In that case, the third metasurface would actually act as a leaky-wave antenna.

\subsection{Simplified Synthesis Based on Leaky-Wave and Guiding-Wave Structures}\label{sec:simp_synth}

The GSTCs-based metasurface synthesis technique~\cite{achouri2014general}, used in the previous section, yields the exact susceptibilities performing the specified transformation. However, the resulting susceptibilities may, in some situations, be difficult to realize. For instance, the metasurface described by the susceptibilities in Figs.~\ref{Fig:Xs} presents spatially varying electric and magnetic losses which may be challenging to implement. Moreover, the generation of the transmitted space wave also requires gain as is evidenced by the positive imaginary parts of the susceptibilities on the right-hand side of Figs.~\ref{Fig:Xs}. For these reasons, we next propose an alternative synthesis method for space-wave to surface-wave transformations performed by the two end metasurfaces in Fig.~\ref{Fig:Compa2}. This method will result in a slightly different design that will be much easier to realize while sacrificing little efficiency.

The waveguiding structure (second metasurface in Fig.~\ref{Fig:Compa2}) will be realized using the susceptibilities found in the previous section since these susceptibilities are exact and easy to realize, as seen in Fig.~\ref{Fig:Xs}. However, the structure will require some optimization to account for deviations from the ideal response due to its non-zero thickness. This may be achieved by following design procedures routinely used in the implementation of slow-wave structures~\cite{lockyear2009microwave,sievenpiper1999high}, as will be discussed thereafter.

\begin{figure}[htbp]
\centering
\subfloat[]{
\includegraphics[width=0.9\linewidth]{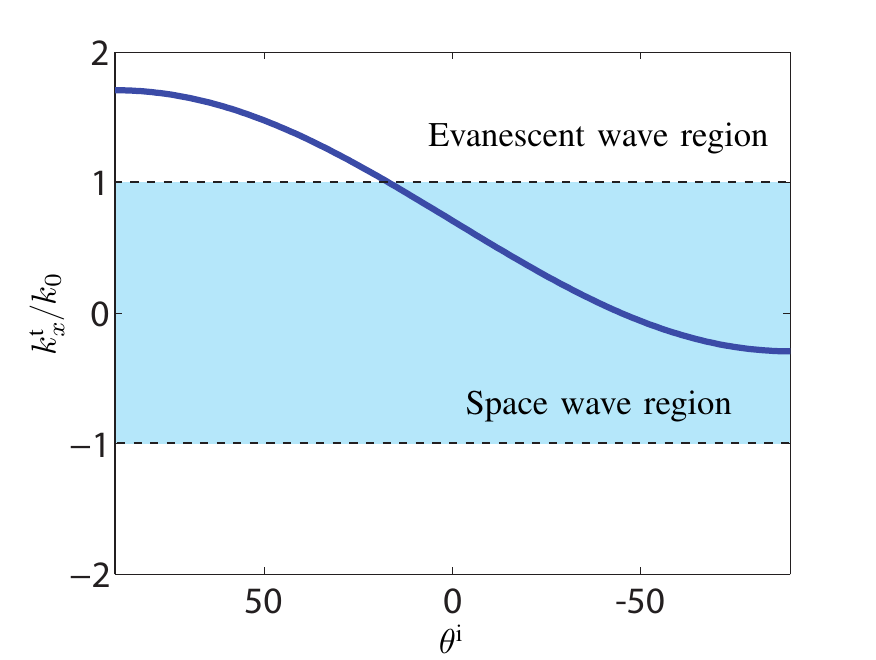}
%\psfragfig*[width=\linewidth]{PGM}{
%\psfrag{b}[][][1]{Evanescent wave region}
%\psfrag{a}[][][1]{Space wave region}
%\psfrag{x}[][][1]{$\theta^\text{i}$}
%\psfrag{y}[][][1]{$k_x^\text{t}/k_0$}}
}
\caption{Normalized transverse wavenumber of the transmitted wave versus incidence angle in the phase-gradient metasurface for the specified angles $\theta^\text{i,spec}=0$ and $\theta^\text{t,spec}=45°$ [Eq.~\eqref{eq:GLR2}].}
\label{Fig:PGM}
\end{figure}

In order to transform the incident space wave into a surface wave with a specific propagation constant along the metasurface, we will use here a simple phase gradient structure. Let us consider the generalized law of refraction~\cite{capasso1}, that can be expressed, using the transverse wavenumber of the incident and refracted waves and the effective wavenumber of the phase gradient structure $K$, as
\begin{equation}
\label{eq:GLR}
k_x^\text{t} = k_x^\text{i} + K,
\end{equation}
where $K=2\pi/P$ with $P$ being the phase-gradient period of the metasurface. This period is designed such that the specified incident wave is refracted at a specified angle, i.e.  $P=\lambda_0/(\sin{\theta^\text{t,spec}}-\sin{\theta^\text{i,spec}})$. From~\eqref{eq:GLR}, we express the normalized transverse wavenumber of the transmitted wave as a function of $K$ and the incidence angle as
\begin{equation}
\label{eq:GLR2}
\frac{k_x^\text{t}}{k_0} = \sin{\theta^\text{i}} + \frac{K}{k_0},
\end{equation}
which allows to determine the transverse wavenumber for \emph{any} incidence angle $\theta^\text{i}$. As an illustration, relation~\eqref{eq:GLR2} is plotted in Fig.~\ref{Fig:PGM} as a function of the incidence angle for the specified angles $\theta^\text{i,spec}=0$ and $\theta^\text{t,spec}=45°$.
The region in blue, where $|k_x^\text{t}/k_0|<1$, corresponds to space-wave modes. Outside of this region, $|k_x^\text{t}/k_0|$ is larger than 1 and the longitudinal wavenumber $k_z^\text{t}=\sqrt{k_0^2 - (k_x^\text{t})^2}$ is therefore imaginary, corresponding to a $z$-evanescent or surface-wave mode. This shows that a simple phase gradient metasurface can be used as a converter between a space wave and a surface wave when the metasurface wavenumber $K$ and the incidence angle $\theta^\text{i}$ are properly chosen~\cite{sun2012gradient,xu2015steering,7497506}.

The three-metasuface system in Fig.~\ref{Fig:Compa2} may therefore be realized as follows. The first metasurface is designed as a phase-gradient metasurface with increasing phase in the $+x$-direction; this positive phase ramp increases the momentum of the incident wave (in the $x$-direction) so as to transform it into a surface wave. The second metasurface is designed to support the propagation of a surface wave with the same wavenumber. Finally, the third metasurface is again designed as a phase-gradient but this time with increasing phase in the $-x$-direction, which reduces the momentum of the surface wave and hence transforms it back into a space wave.

\section{Experimental Demonstration:\\ The ``Electromagnetic Periscope''}
\label{sec:4}
\subsection{Realization}

For the realization of the metasurface system, and particularly the realization of the space-wave - surface-wave converters (metasurfaces 1 and~3), we use, for simplicity, the approximate synthesis technique presented in Sec.~\ref{sec:simp_synth}, rather than the exact but more problematic technique based on GSTCs presented in Sec.~\ref{sec:Ex_Synthesis_GSTCs}.

\begin{figure}[htbp]
\centering
\includegraphics[width=1\linewidth]{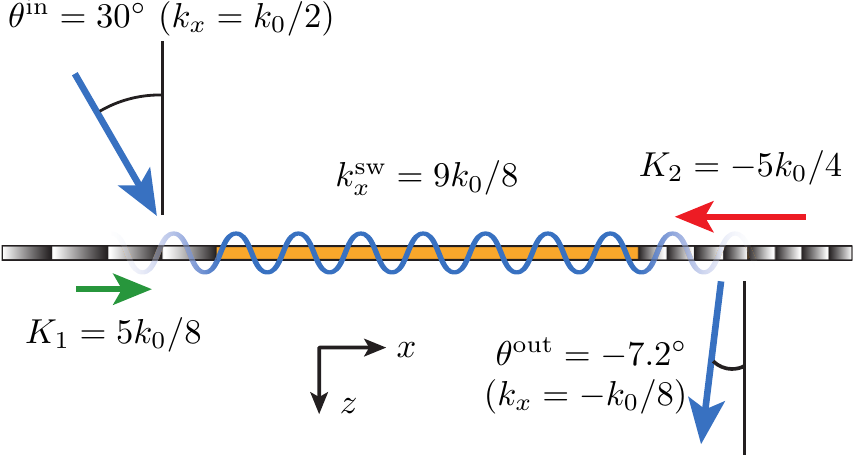}
%\psfragfig*[width=\linewidth]{diagram}{
%\psfrag{a}[][][1]{$\theta^\text{in}=30^\circ~(k_x =k_0/2)$ }
%\psfrag{b}[][][1]{$k_x^\text{sw}=9k_0/8$}
%\psfrag{c}[][][1]{\begin{tabular}{@{}l@{}}
%   $~\theta^\text{out}=-7.2^\circ$\\ $(k_x =-k_0/8)$
%\end{tabular}}
%\psfrag{d}[][][1]{$K_1 = 5k_0/8$}
%\psfrag{e}[][][1]{$K_2 = -5k_0/4$}}
\caption{Schematic representation of the electromagnetic periscope metasurface system.}
\label{Fig:Schem}
\end{figure}

A schematic of the metasurface system is presented in Fig.~\ref{Fig:Schem}. The figure shows the conceptual operation of the structure with momentum ``push''~($K_1$) and momentum ``pull'' ($K_2$) induced by the first and last metasurfaces, respectively, and the surface-wave guidance in the middle metasurface. We shall now design the metasurface system for the following specifications: input angle $\theta^\text{in}=30^\circ$ and output angle $\theta^\text{out}=-7.2^\circ$.

The first phase-gradient metasurface, transforming the input $p$-polarized space wave into a surface wave, will be implemented with a supercell of 8 unit cells of size $\lambda_0/5$ with transmission phases ranging from 0 to $2\pi$. The corresponding metasurface wavenumber is $K_1=2\pi/P_1=2\pi/(8\lambda_0/5)=5k_0/8$. Then, for the specified input wave of $\theta^\text{in}=30^\circ$, corresponding to $k_x^\text{in}=k_0/2$, one finds, using~\eqref{eq:GLR} with $k_x^\text{i}=k_x^\text{in}$ and $K=K_1$, the surface-wave wavenumber to be $k_x^\text{sw}=k_x^\text{t}=9k_0/8$, which corresponds to the $x$-wavenumber across the entire metasurface system.

Upon this basis, the third metasurface is designed as follows. The output angle of $\theta^\text{out} = -7.2^\circ$ corresponds to $k_x^\text{out}=-k_0/8$. We apply~\eqref{eq:GLR} with $k_x^\text{i}=k_x^\text{sw}=9k_0/8$ and $k_x^\text{t}=k_x^\text{out}=-k_0/8$, which yields $K_2=-5k_0/4$. Since $|K_2/K_1|=2$, $P_2=P_1/2$, and hence, still assuming $\lambda_0/5$ unit cells, the supercell includes now 4 unit cells. This metasurface may for instance be identical to the first metasurface where every two unit-cell rows have been removed.

The $p$-polarization surface-wave guiding structure, in the middle of the metasurface system, may also be realized as a metasurface, for compatibility with its phase-gradient neighbours, instead of as a traditional waveguiding structure. In addition, to allow the $s$-polarization generalized refraction operation depicted in Fig.~\ref{Fig:concept}, this structure must be completely transparent, and could therefore not be implemented in the form of a conventional waveguide. As explained in~\ref{sec:simp_synth}, the metasurface is designed using the $p$-polarization susceptibilities already found with the exact synthesis [Eqs.~\eqref{eq:chi_diag1}], namely $\chi_\text{ee}^{xx}=2j/k_z^\text{sw}$ and $\chi_\text{mm}^{yy}=2jk_z^\text{sw}$, where $k_z^\text{sw}=\sqrt{k_0^2-(k_x^\text{sw})^2}$ with the value $k_x^\text{sw}=9k_0/8$ found above. In the current design, we consider the particular case of $s$-polarization normal transmission, leading to global uniformity.
\\
\begin{figure}[htbp]
\centering
\includegraphics[width=1\linewidth]{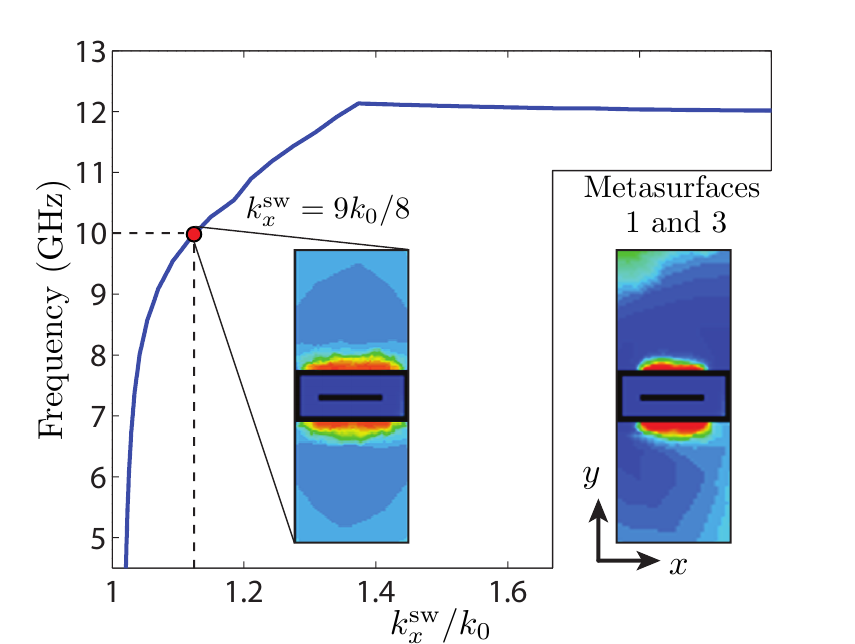}
%\psfragfig*[width=\linewidth]{dispersion}{
%\psfrag{a}[][][0.9]{$k_x^\text{sw}=9k_0/8$}
%\psfrag{b}[][][0.9]{\begin{tabular}{@{}l@{}}
%   Metasurfaces\\ ~~~~1 and 3
%\end{tabular}}
%\psfrag{c}[][][1]{$x$}
%\psfrag{d}[][][1]{$y$}
%\psfrag{x}[][][1]{$k_x^\text{sw}/k_0$}
%\psfrag{y}[][][1]{Frequency (GHz)}}
\caption{Dispersion curve and magnetic field distribution (absolute value at 10~GHz) for the fundamental mode of the waveguiding metasurface. The separate inset represents the excited fields in the surrounding phase-gradient metasurfaces (also at 10~GHz).}
\label{Fig:PGM_disp}
\end{figure}

The overall metasurface system, composed of the three juxtaposed metasurfaces, is implemented as a multilayer structure with three metallization layers and two dielectric spacers. The overall thickness of the structure is $\lambda_0/10$. In each metasurface, the unit cell has a transverse size of $\lambda_0/5\times\lambda_0/5$ and includes in each layer a metallic scatterer in the form of a Jerusalem cross with specific geometric parameters~\cite{GrbicCascaded2013,PhysRevApplied.2.044011,AchouriEPJAM,achouri2016metasurface}.

The exact dimensions of the Jerusalem crosses are found by numerical simulations using a commercial software. Each unit cell is simulated individually assuming periodic boundary conditions as approximate boundaries for smoothly varying patterns. The resulting scattering parameters, obtained from the simulations, are optimized by varying the dimensions of the crosses until the expected response is achieved~\cite{GrbicCascaded2013,PhysRevApplied.2.044011,AchouriEPJAM,achouri2016metasurface}. For the two phase-gradient metasurfaces, the scattering parameters of each unit cell are assumed to simply consist of a phase transmission coefficient, $T=e^{j\phi}$, where the phase shift $\phi$ depends on the unit cell position within the supercell. For the waveguiding metasurface, the susceptibilities given above are first converted into scattering parameters following the procedure given in~\cite{achouri2014general}. Because this metasurface is uniform (as seen by a $p$-polarized wave), in contrast to the phase-gradient metasurfaces, only one unit cell has to be designed. Once the dimensions of the Jerusalem crosses corresponding to the susceptibilities have been found, the unit cell is optimized using an eigenmode solver with the goal to achieve a wavenumber of $k_x^\text{sw}=9k_0/8$ at the operation frequency set here to $f=10$~GHz. The dispersion curve for the fundamental mode of the optimized waveguiding structure is plotted in Fig.~\ref{Fig:PGM_disp}. Note that the horizontal axis represents the $x$-wavenumber normalized to the free-space wavenumber, so that the figure shows only the slow-wave region ($k_x^\text{sw}/k_0>1$). Comparing the two insets in the figure shows that the field distribution of this fundamental mode is essentially identical, and hence compatible, with the field distributions of the two phase-gradient metasurfaces. Since the metasurfaces have been in addition designed to all exhibit the same polarization and wavenumber, it may be inferred that the coupling between them is maximized, as desired.

\begin{figure}[htbp]
\centering
\includegraphics[width=1\linewidth]{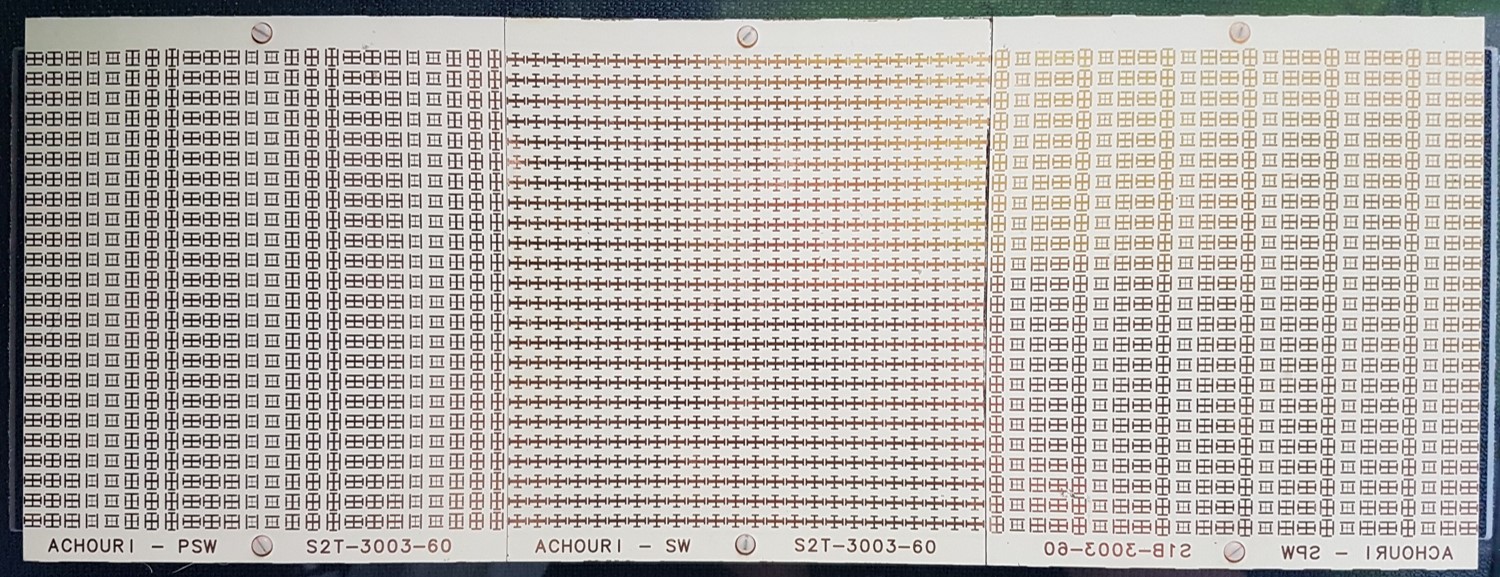}
\caption{Fabricated metasurface system corresponding to Fig.~\ref{Fig:concept}. The metasurfaces from the left to the right perform the following operations on the $p$-polarized wave: space-wave to surface-wave transformation, surface-wave propagation, and surface-wave to space-wave transformation. At the same time, the central metasurface is perfectly transparent to $s$-polarized waves. The difference between the phase-gradients of the two end metasurfaces is clearly visible.}
\label{Fig:MS}
\end{figure}

The realized metasurface system is shown in Fig.~\ref{Fig:MS}. Due to limitation of our fabrication process, the three metasurfaces have been realized separately rather than as a single entity and have then been screwed to a plastic frame (at the back and hence not visible in Fig.~\ref{Fig:MS}) to form the overall metasurface system. Each metasurface is made of $24\times24$ unit cells, corresponding to a size of $4.8\lambda_0\times4.8\lambda_0$. The dimensions of the system are 45~cm$\times$15~cm$\times3$~mm.

\subsection{Experiment}

\begin{figure}[htbp]
\centering
\includegraphics[width=1\linewidth]{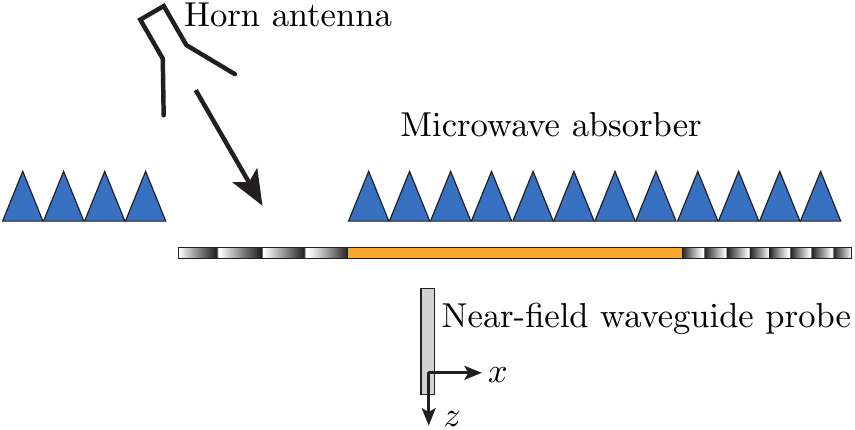}
%\psfragfig*[width=\linewidth]{Setup}{
%\psfrag{a}[][][1]{Horn antenna}
%\psfrag{b}[][][1]{Microwave absorber}
%\psfrag{c}[][][1]{Near-field waveguide probe}
%\psfrag{x}[][][1]{$x$}
%\psfrag{z}[][][1]{$z$}}
\caption{Side view of the metasurface system measurement setup.}
\label{Fig:Setup}
\end{figure}

The measurement of the metasurface system was performed using the experimental setup depicted in Fig.~\ref{Fig:Setup}. The input side of the metasurface system is covered everywhere by absorbers except for a small aperture allowing the illumination of the first metasurface on the left. A high-gain X-band horn antenna illuminates the structure at the input side while a waveguide probe scans the metasurface system at the output side in the near-field region. The near-field is measured in the middle of the metasurface system in Fig.~\ref{Fig:MS} along the $x$-direction. The measured near-field will be first Fourier-transformed to compute the spatial ($k$-domain) spectrum, and hence identify the modes excited at the output side of the system, and next propagated in the $xz$-plane by the angular spectrum technique~\cite{novotny2012principles}, so as to verify the periscope operation of the system.

\begin{figure}[htbp]
\centering
\subfloat[]{
\includegraphics[width=0.9\linewidth]{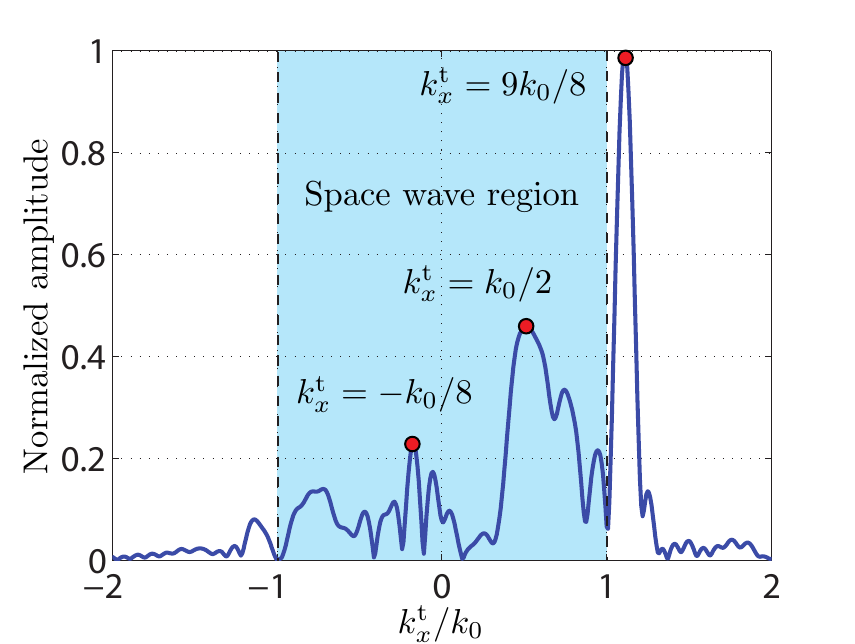}
%\psfragfig*[width=\linewidth]{all_MS}{
%\psfrag{a}[][][1]{$k_x^\text{t}=9k_0/8$}
%\psfrag{b}[][][1]{$k_x^\text{t}=k_0/2$}
%\psfrag{d}[][][1]{$k_x^\text{t}=-k_0/8$}
%\psfrag{c}[][][1]{Space wave region}
%\psfrag{x}[][][1]{$k_x^\text{t}/k_0$}
%\psfrag{y}[][][1]{Normalized amplitude}}
\label{Fig:MeasMS1}}\\
\subfloat[]{
\includegraphics[width=0.9\linewidth]{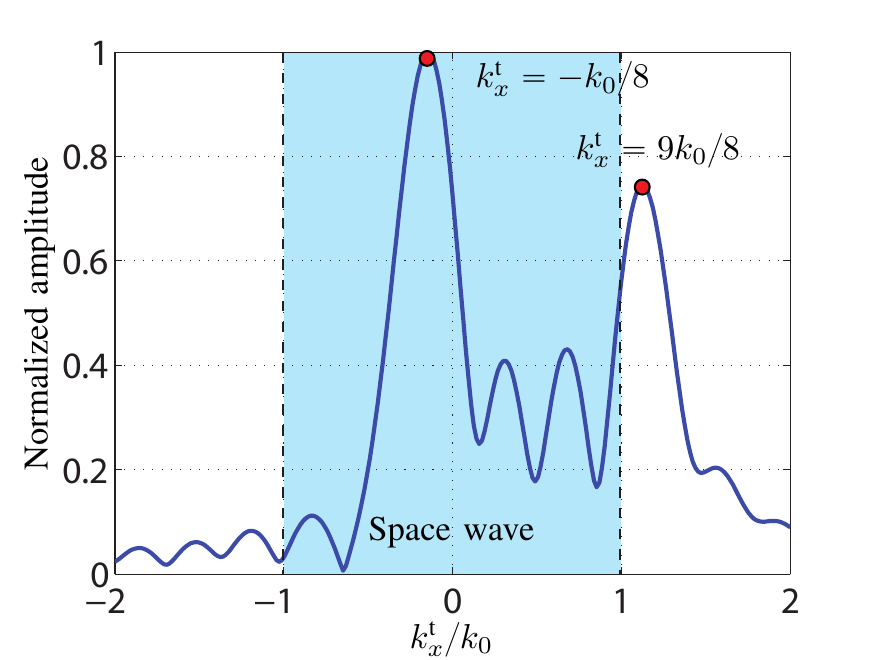}
%\psfragfig*[width=\linewidth]{SPW}{
%\psfrag{a}[][][1]{$k_x^\text{t}=9k_0/8$}
%\psfrag{b}[][][1]{$k_x^\text{t}=-k_0/8$}
%\psfrag{c}[][][1]{Space wave}
%\psfrag{x}[][][1]{$k_x^\text{t}/k_0$}
%\psfrag{y}[][][1]{Normalized amplitude}}
\label{Fig:MeasMS2}}
\caption{Normalized $x$-Fourier transform ($k_x$-domain) of the output near-field measured along the $x$-direction at 1~cm from the metasurface in the $z$-direction (Fig.~\ref{Fig:Setup}). (a)~Scanning across the entire metasurface system. (b)~Scanning only across the third metasurface. The regions highlighted in blue correspond to the radiation region.}
\label{Fig:MeasMS}
\end{figure}

The modes excited along the overall structure, as the probe scans the entire $x-$dimension of the system, are revealed in Fig.~\ref{Fig:MeasMS1}, which plots the normalized $x$-Fourier transform of the output near-field measured along the $x$-direction using the setup of Fig.~\ref{Fig:Setup}. The mode excited at the output of the metasurface system with the highest amplitude is a surface wave of wavenumber $k_x^\text{sw}=9k_0/8$ corresponding to the wavenumber of the specified surface-wave mode. The reason why this mode is dominant is because it is excited along the entire structure, being first generated on the first metasurface, next guided by the second one and eventually radiated by the third one. The mode excited with the next higher intensity is the space-wave mode at $k_x^\text{t}=k_0/2$, which corresponds to the input wave impinging the metasurface at $\theta^\text{in}=30^\circ$. The third largest peak lies in the negative side of the horizontal axis and corresponds to the specified transmitted space wave with wavenumber $k_x^\text{t}=-k_0/8$ generated by the third metasurface.

Figure~\ref{Fig:MeasMS2} shows the modes excited only at the output side of the third metasurface, when the near-field probe scans only on that part of the system. As expected, the two strongest modes correspond to the specified transmitted space wave with $k_x^\text{t}=-k_0/8$ and the specified surface wave with $k_x^\text{t}=9k_0/8$.

\begin{figure}[htbp]
\centering
\includegraphics[width=1\linewidth]{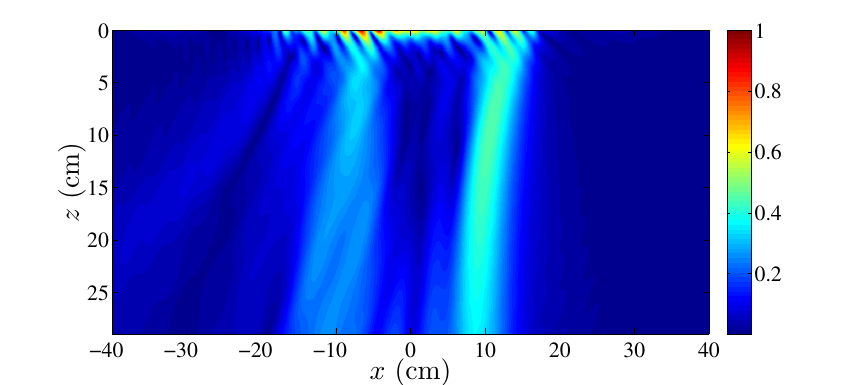}
%\psfragfig*[width=\linewidth]{propag}{
%\psfrag{x}[][][0.8]{$x$ (cm)}
%\psfrag{z}[][][0.8]{$z$ (cm)}}
\caption{Absolute value of the transmitted electric field ($E_x$ component) obtained by angular spectrum propagation. The metasurface system is at $z=0$ and extends from $x=-22.5$~cm to $x=22.5$~cm.}
\label{Fig:Propag}
\end{figure}

Next, we compute the field scattered from the metasurface system by applying the angular spectrum propagation technique~\cite{novotny2012principles} to the near-field measured along the entire structure. To clearly see the propagation of the expected transmitted space wave with $k_x^\text{t}=-k_0/8$, we ignore the contribution of the input wave, which generates important spurious scattering, as is visible in Fig.~\ref{Fig:MeasMS1} around $k_x^\text{t}/k_0=0.5$. This is achieved by first taking the Fourier transform of the near-field, yielding the data in Fig.~\ref{Fig:MeasMS1}, and next setting to zero all the modes excited in the region $0.2< k_x^\text{t}/k_0 <0.8$ in Fig.~\ref{Fig:MeasMS1} to remove the contributions from the input wave. Then, the field is propagated along the $z$-direction following the usual procedure of the angular spectrum propagation technique. The resulting scattered field is plotted in Fig.~\ref{Fig:Propag}, where the metasurface system lies at $z=0$ and extends from $x=-22.5$~cm to $x=22.5$~cm. In this figure, we can see the presence of a strong surface wave near the structure close to $z=0$. In the region around $x=-10$~cm, we see some scattering which is due to the discontinuity between adjacent metasurfaces. In the region around $x=10$~cm, we see a beam emerging from the metasurface system and being deflected towards the left. This beam corresponds to the specified transmitted space wave with $\theta^\text{out}=-7.2^\circ$.

\begin{figure}[htbp]
\centering
\includegraphics[width=1\linewidth]{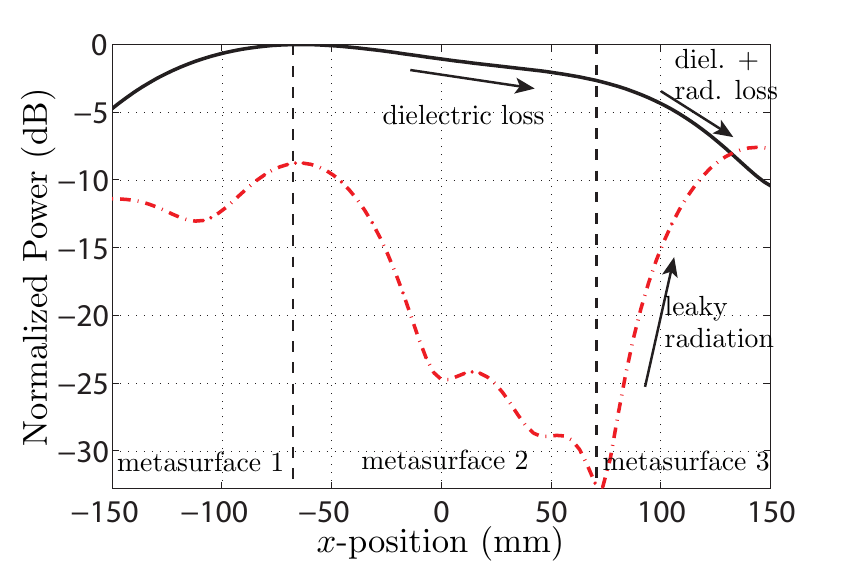}
%\psfragfig*[width=\linewidth]{Amp2}{
%\psfrag{a}[][][0.8]{metasurface 1}
%\psfrag{b}[][][0.8]{metasurface 2}
%\psfrag{c}[][][0.8]{metasurface 3}
%\psfrag{d}[][][0.8]{dielectric loss}
%\psfrag{e}[][][0.8]{\begin{tabular}{@{}l@{}}
%   diel. +\\ rad. loss
%\end{tabular}}
%\psfrag{f}[][][0.8]{\begin{tabular}{@{}l@{}}
%   leaky\\radiation
%\end{tabular}}
%\psfrag{x}[][][1]{$x$-position (mm)}
%\psfrag{y}[][][1]{Normalized Power (dB)}}
\caption{Normalized power distribution of the surface-wave mode (solid black line) and of the transmitted wave (dashed red line) over the metasurface system. The two vertical dashed black lines indicate the separation between the three metasurfaces. }
\label{Fig:Amp}
\end{figure}

In order to better understand the result shown in Fig.~\ref{Fig:Propag}, we next analyze the spatial power distributions of the surface wave ($k_x^\text{sw} = 9k_0/8$) and of the transmitted space wave ($k_x^\text{out} = -k_0/8$) along the metasurface system. From the data plotted in Figs.~\ref{Fig:MeasMS}, it is possible to extract the power distribution of the different modes over the metasurface system. This is achieved by first isolating the modes of interest in the data of Fig.~\ref{Fig:MeasMS1} by setting to zero everything except the relevant regions (appropriate peaks) -- for example leaving only the peak centered at $k_x^\text{t} = 9k_0/8$ to isolate the surface wave -- and then taking the inverse Fourier transform to generate the spatial distribution of the mode. The results are presented in Fig.~\ref{Fig:Amp}, where the power distribution of the surface wave is represented by the solid black line and that of the space wave by the dashed red line.

As one moves along the $x$-axis, the  power distribution  of the surface wave (solid black curve) first increases, following the power distribution of the exciting horn antenna, which points at the junction between the first and second metasurfaces. At this point, it reaches a corresponding maximum. Then, it decreases as the wave propagates along the waveguiding metasurface while experiencing metallic and dielectric dissipation losses. Finally, it further decreases on the third metasurface due to combined dissipation and radiation losses.

The power level of the transmitted space wave (dash red curve) is relatively high at the junction between the first and second metasurfaces, which is explained by spurious scattering of the incident wave at this discontinuity, similarly to the undesired scattering apparent in Figs.~\ref{Fig:Sim2} and~\ref{Fig:Sim3}. Then, this power rapidly decreases along the second metasurface, as expected from the fact that this surface does not radiate. Along the third metasurface, the power of the space wave progressively increases as it is progressively generated in terms of leaky-wave radiation by the interaction between the surface wave and the phase-gradient of the metasurface.

The experimental results presented above are in perfect agreement with the expected response of the metasurface system, with the exception of a relatively low efficiency of about $10\%$. This low efficiency is due to a combination of effects that include surface-wave dissipation loss, scattering at each of the two metasurface discontinuities, the limited coupling of the incident wave which is effectively converted to a surface wave, and the imperfect conversion between space wave and surface wave (and vice-versa) due to the simplified synthesis technique used for the implementation of the phase-gradient metasurfaces. Several of these issues may be addressed by further optimization.

\section{Other Potential Applications}
\label{sec:5}

The concept of space-wave via surface-wave routing may lead to a diversity of other potential applications. As an illustration, we will discuss two of them in this section.

\subsection{Compact Beam Expander}

An optical beam expander is a device that is used in telescopes or microscopes: it increases (or decreases) the lateral size of the incoming beam. The simplest way to realize such a device is to cascade two thin lenses of different focal lengths. We propose here an alternative beam expanding system, based on the concept of space-wave via surface-wave routing. Compared to the lens system, this routing system presents two significant advantages. First, it uses a single (composite) metasurface instead of two lenses. Second, in contrast to the lens system, it does not require any separation distance, where such a distance at optical frequencies represents several thousands of wavelengths, and hence it leads a very compact system.

\begin{figure}[htbp]
\centering
\subfloat[]{
\includegraphics[width=1\linewidth]{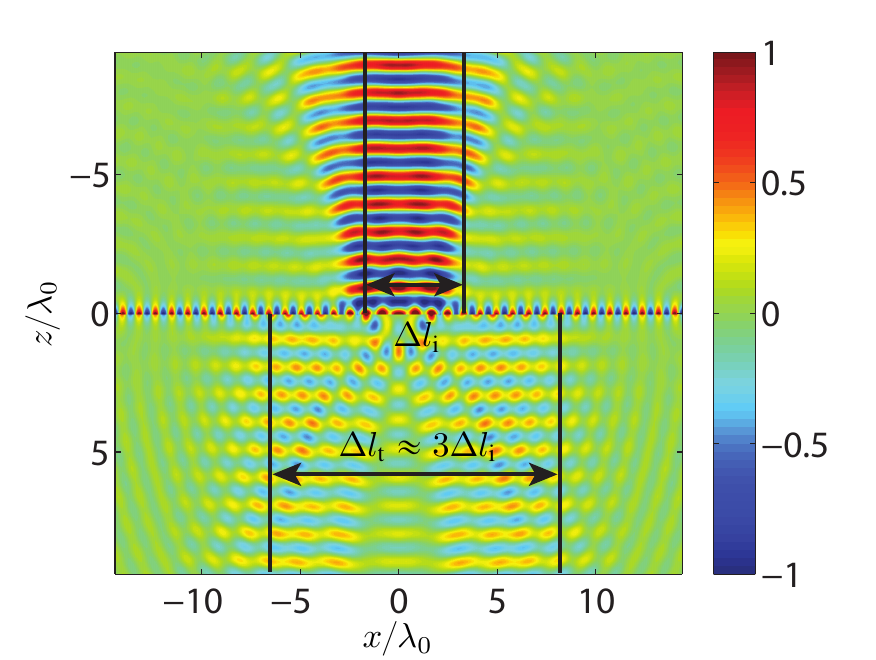}
%\psfragfig*[width=\linewidth]{DBE}{
%\psfrag{a}[][][1]{$\Delta l_\text{i}$}
%\psfrag{b}[][][1]{$\Delta l_\text{t}\approx 3\Delta l_\text{i}$}
%\psfrag{x}[][][1]{$x/\lambda_0$}
%\psfrag{z}[][][1]{$z/\lambda_0$}}
\label{Fig:BE1}}\\
\subfloat[]{
\includegraphics[width=1\linewidth]{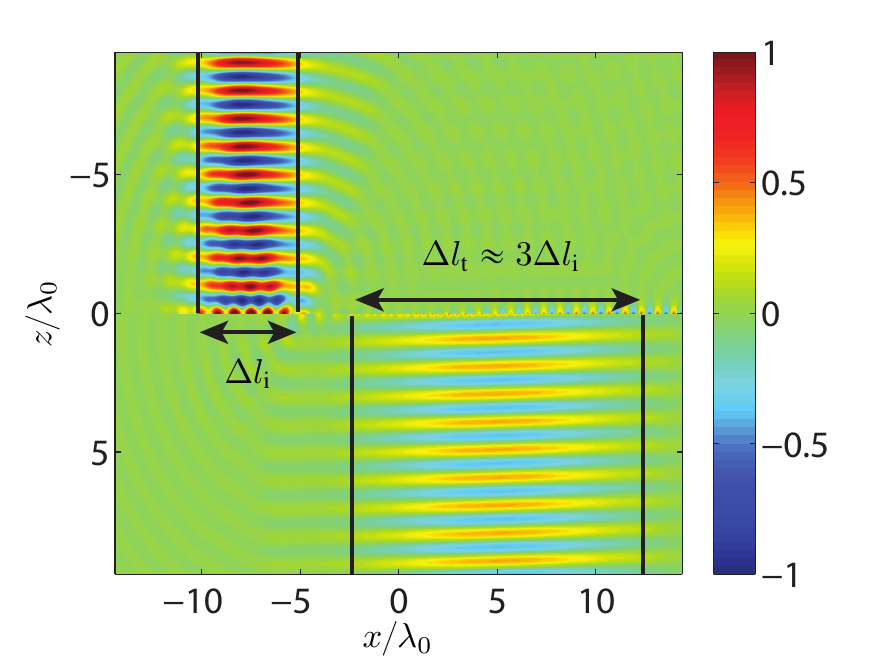}
%\psfragfig*[width=\linewidth]{OBE}{
%\psfrag{a}[][][1]{$\Delta l_\text{i}$}
%\psfrag{b}[][][1]{$\Delta l_\text{t}\approx 3\Delta l_\text{i}$}
%\psfrag{x}[][][1]{$x/\lambda_0$}
%\psfrag{z}[][][1]{$z/\lambda_0$}}
\label{Fig:BE2}}
\caption{FDFD simulation of a beam expander with (a)~direct transformation and (b)~offset transformation. The metasurface system is designed to increase the beamwidth of the incident wave by a factor 3.}
\label{Fig:BE}
\end{figure}

We present here two different beam expander designs, both increasing the beam width by a factor $3$. One performs a direct conversion (without any lateral shift) while the other one performs an offset (laterally shifted) beam expansion. The direct beam expander is made of three metasurfaces, the middle one transforming the incident beam into two contra-propagating surface waves that are then both transformed back into space waves by the two end metasurfaces. The simulation showing this direct expansion is presented in Fig.~\ref{Fig:BE1}. As may be seen in this plot, the presence of the two metasurface discontinuities induces important spurious scattering.

The simulation of the offset beam expander is shown in Fig.~\ref{Fig:BE2}. The system is identical to that of Fig.~\ref{Fig:Sim3} except that both the incident and transmitted angles are now normal to the surface. For the two structures in Figs.~\ref{Fig:BE}, the beam expansion of the transmitted wave is about three times that of the incident wave. Consequently, the amplitude of the transmitted wave is also three times less.

\subsection{Multi-wave Refractor}

\begin{figure}[htbp]
\centering
\includegraphics[width=0.9\linewidth]{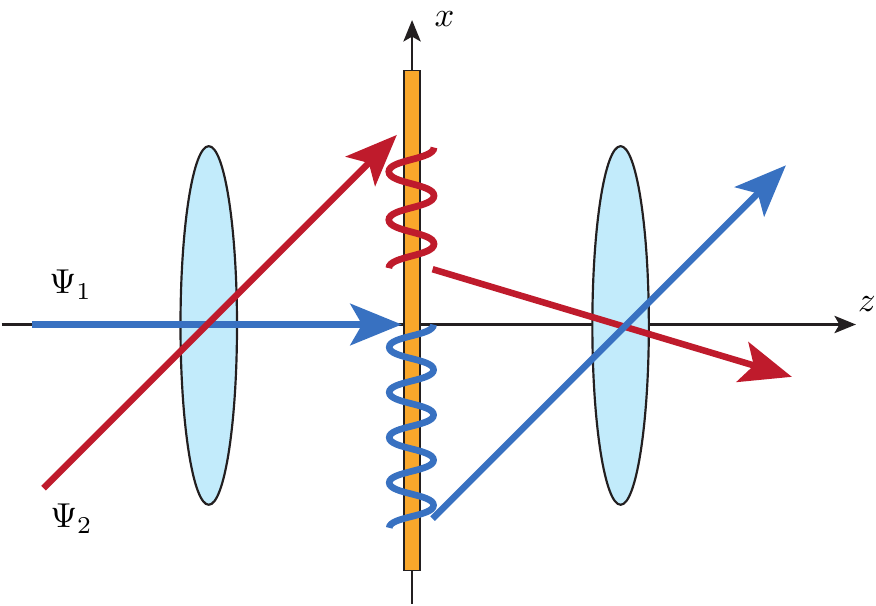}
%\psfragfig*[width=\linewidth]{MultRefrac}{
%\psfrag{a}[][][1]{$\Psi_1$}
%\psfrag{b}[][][1]{$\Psi_2$}
%\psfrag{x}[][][1]{$x$}
%\psfrag{z}[][][1]{$z$}}
\caption{Multi-wave refractor consisting of a 4$-f$ system, with 2 routing metasurface systems in its Fourier plane, which refracts the input waves $\Psi_1$ and $\Psi_2$ into different directions.}
\label{Fig:Refractor1}
\end{figure}

The capability to route beams via surface waves may also be leveraged to implement a multi-wave refractor, i.e. system performing several refractive transformations with a single metasurface system, in contrast to a conventional metasurface that can only perform two independent refraction transformations, one for an $x$-polarized wave and one for a $y$-polarized wave (or up to 4 refractions by leveraging nonreciprocity and making use of gain and loss, as discussed in~\cite{7428120}). The proposed system is realized by inserting a metasurface at the Fourier plane of an optical 4-$f$ system. A 4-$f$ system is generally used as a spatial filter where a mask is placed at the Fourier plane to filter out certain spatial components of the incident wave~\cite{saleh2007fundamentals}. Here, the metasurface placed at the Fourier plane is not used to filter out spatial components but, instead, to shift the spatial components of the incident waves to another region of the plane, which effectively changes the direction of propagation of the transmitted waves. The concept is depicted in Fig.~\ref{Fig:Refractor1}, where two input beams, $\Psi_1$ and $\Psi_2$, are transformed in terms of their spectral contents in the 4-$f$ system.
\begin{figure}[htbp]
\centering
\includegraphics[width=0.8\linewidth]{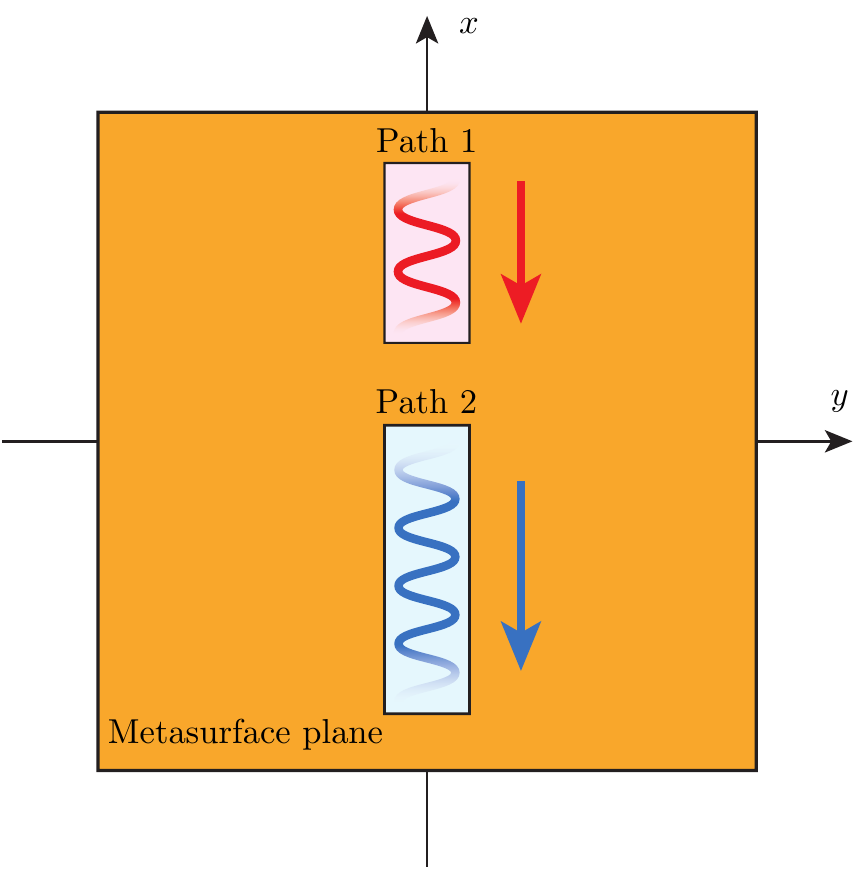}
%\psfragfig*[width=\linewidth]{MetaPath}{
%\psfrag{a}[][][1]{Metasurface plane}
%\psfrag{b}[][][1]{Path 1}
%\psfrag{c}[][][1]{Path 2}
%\psfrag{x}[][][1]{$y$}
%\psfrag{y}[][][1]{$x$}}
\caption{Representations of the metasurface at the Fourier plane of Fig.~\ref{Fig:Refractor1} with the two optical routes shifting the waves $\Psi_1$ and $\Psi_2$ to different locations in the Fourier plane.}
\label{Fig:Refractor2}
\end{figure}
The first lens focalizes the two beams at different locations in the Fourier plane, where a metasurface system is placed. This metasurface system consists of two ``optical routes'', as shown in Fig.~\ref{Fig:Refractor2}, each composed of three different metasurfaces successively transforming the incident space wave into a surface wave, guiding this surface wave along the the Fourier plane to the appropriate $(k_x,k_y)$ point, and transforming it back into a space wave in the desired direction. In this example, the two beams have been shifted along the $-x$-direction in the Fourier plane. Their respective momenta along $x$ have therefore been decreased. Consequently, the two beams exit the system, collimated by the second lens, with transmission angle depending on the points to which they have been shifted in the Fourier plane. Such a metasurface system might be populated with several additional ``optical routes'' so as to achieve even more refraction transformations.

\section{Conclusion}\label{sec:6}

In this work, we have introduced the concept of space-wave via surface-wave routing system composed of several juxtaposed metasurfaces. We have presented two synthesis techniques to design such a routing system. One is exact and most efficient but also complex to realize in practice due to the presence of spatially varying electric and magnetic losses. The other one is an alternative simplified approach that consists in using phase-gradient metasurfaces, to generate the surface wave, and on dispersion engineering, to guide the surface wave along the structure. This alternative synthesis technique is based on the generalized law of refraction, which is approximate, and therefore leads to a less efficient structure, due to undesired scattering, but offers the advantage of being easier to realize.

As a proof of concept, we have presented a metasurface system acting as an ``electromagnetic periscope''. This system spatially shifts an incident beam impinging the structure under a given angle and then reradiates it under a small angle. This structure produces the expected result, but suffers from a relatively low efficiency, which may be improved by further optimization. To illustrate the capabilities of the proposed concept, we have also presented two other potential applications, namely a compact beam expander and a multi-wave refractor that may be used as a spatial coupler between multiple inputs and outputs.

\section*{Acknowledgment}

This work was accomplished in the framework of the Collaborative Research and Development Project CRDPJ 478303-14 of the Natural Sciences and Engineering Research Council of Canada (NSERC) in partnership with the company Metamaterial Technology Inc.

\bibliography{LIB}

\end{document}